\documentclass[reprint,twocolumn,showpacs,preprintnumbers,showkeys,amsmath,amssymb,pra,10pt]{revtex4-1}

\usepackage[T1]{fontenc}
\usepackage{graphicx}
\usepackage{epstopdf}
\usepackage{dcolumn}
\usepackage{bm}

\newcommand{\szerokosc}{0.45}
\usepackage{hyperref}

\begin{document}
\preprint{Submitted to: ACTA PHYSICA POLONICA A}

\title{Some properties of two dimensional extended repulsive Hubbard model\\ with intersite magnetic interactions --- a~Monte Carlo study}

\author{Szymon Murawski}%
\author{Konrad Jerzy Kapcia}%
    \email{corresponding author; e-mail: konrad.kapcia@amu.edu.pl}
\author{Grzegorz Paw\l{}owski}%
\author{Stanis\l{}aw Robaszkiewicz}%
\affiliation{Electron States of Solids Division, Faculty of Physics, Adam Mickiewicz University in Pozna\'n, Umultowska 85, 61-614 Pozna\'n, Poland
}

\date{October 6, 2014}

\begin{abstract}
In this paper the two dimensional extended Hubbard model  with intersite magnetic Ising-like interaction in the atomic limit is analyzed by means of the classical Monte Carlo method in the grand canonical ensemble. Such an~effective simple model could describe behavior of insulating (anti)ferromagnets. In the model considered  the Coulomb interaction ($U$) is on-site and the magnetic interactions in $z$-direction (\mbox{$J>0$}, antiferromagnetic) are restricted to nearest-neighbors. Simulations of the model have been performed on a~square lattice consisting of \mbox{$N=L\times L=400$} sites (\mbox{$L=20$}) 
in order to obtain the full phase diagram for \mbox{$U/(4J)=1$}.
Results obtained for on-site repulsion (\mbox{$U>0$}) show that, apart from homogeneous non-ordered (NO) and ordered magnetic (antiferromagnetic, AF) phases, there is also a region of phase separation (PS: AF/NO) occurrence. We present a~phase diagram  as well as some thermodynamic properties of the model for the case of \mbox{$U/(4J)=1$} (and arbitrary chemical potential and arbitrary electron concentration). The \mbox{AF--NO} transition can be second-order as well as first-order and the tricritical point occurs on the diagram.
\end{abstract}




\maketitle

\section{Introduction}

Since its introduction in 1963 \cite{H1963} the Hubbard model has found applications in many various systems. Despite half-century research on this model it still holds a number of open questions. This report focuses on the atomic limit of the extended Hubbard model, in which we restrict ourselves to the case of the zero-bandwidth limit (\mbox{$t=0$}).
with added magnetic interactions of the Ising-type between electrons. Such a~simple model can be used for describing behavior of insulating magnets.
The hamiltonian of the discussed model has the following form:
\begin{equation*}
\hat{H} = U\sum_i{\hat{n}_{i\uparrow}\hat{n}_{i\downarrow}} + 2J\sum_{\langle i,j\rangle}{\hat{s}^z_{i}\hat{s}^z_{j}} - \mu\sum_{i}{\hat{n}_{i}}
\end{equation*}
where $U$~is the on-site density interaction,
$J$~is $z$-component of the intersite magnetic exchange interaction,
$\mu$ is chemical potential,
and $\sum_{\left\langle i,j\right\rangle}$ restricts the summation to nearest-neighbor sites.
$n_i=\hat{n}_{i\uparrow}+\hat{n}_{i\downarrow}$ is total electron number on site $i$,
whereas $\hat{s}^z_i=(1/2)(\hat{n}_{i\uparrow}-\hat{n}_{i\downarrow})$ is $z$-component of total spin at $i$ site.
\mbox{$\hat{n}_{i\sigma}=\hat{c}^{+}_{i\sigma}\hat{c}_{i\sigma}$} is electron number with spin $\sigma$ on site $i$, where $\hat{c}^{+}_{i\sigma}$ and $\hat{c}_{i\sigma}$ denote the creation and annihilation operators, respectively, of an electron with spin \mbox{$\sigma=\uparrow,\downarrow$} at the site $i$. The electron concentration $n$ is defined as \mbox{$n=(1/N)\sum_i \langle \hat{n}_i \rangle$}, where $N$ is the total number of sites.

The rigorous ground state results for this model have been found in the case of a~\mbox{$d=1$} chain \cite{MPS2012,MPS2013} and \mbox{$2\leq d< +\infty$} case \cite{BS1986,J1994}. The exact results for finite temperature have been also obtained \cite{MPS2013} for \mbox{$d=1$} chain (an absence of long-range order at \mbox{$T>0$}).
Within the variational approach the model has been analyzed for half-filing (\mbox{$n=1$}) \cite{R1975,R1979} as well as for arbitrary electron concentration \mbox{$0\leq n \leq 2$} \cite{KKR2010,KKR2012} (these results are rigorous in the limit of infinite dimensions \mbox{$d\rightarrow+\infty$}).
Our preliminary Monte Carlo (MC) results have been presented in \cite{MKPR2012} for strong on-site repulsion (\mbox{$U/4J=1,10$} and \mbox{$L=10$}). In this paper we investigate in details the phase diagram and thermodynamic properties of the model for arbitrary electron concentration \mbox{$n\leq1$} and arbitrary chemical potential \mbox{$\bar{\mu}\leq 0$} (\mbox{$\bar{\mu}=\mu-U/2$}) in the whole range of temperatures for a specific repulsive value of the on-site interaction parameter \mbox{$U/(4J)=1$} (and \mbox{$L=20$}). The corresponding results for \mbox{$n>1$} and \mbox{$\bar{\mu}>0$} are obvious because of the electron-hole symmetry of the model on alternate lattices.

\section{Monte Carlo simulation details}

The Monte Carlo simulations for the model described above have been done at finite temperatures \mbox{$T>0$} using grand canonical ensemble on two dimensional (\mbox{$d=2$})  square (SQ) lattice with number of neighbors \mbox{$z=4$}. One could depict such approach as adsorption of electron gas on a lattice.
Our simulations  use a~local update method \cite{H1990} determined by elementary ``runs'':
(i)~a~particle can transfer to another site,
(ii)~it can be adsorbed on a~lattice or
(iii)~it can be removed from the lattice.
These ``runs'' are usually called \textit{move}, \textit{create} and \textit{destroy} of particle procedures in MC simulations. Every Monte Carlo step (MCS) consists of each of these ``runs'' performed \mbox{$N=L \times L$} times. In our simulations the number of MCS is \mbox{$10^6$} with a quarter of them being spent on thermalization, which is necessary  to avoid results  heavily influenced by the starting point of the simulations.
There are also many cluster update algorithms, but due to chemical potential term in the hamiltonian they cannot be implemented here.
More details on the simulation method can be found in \cite{P2006}.

Simulations provide data for temperature and chemical potential dependencies of various thermodynamic variables. The variables of the particular interest are:
staggered magnetization \mbox{$m_Q=(m_A-m_B)/2$},
magnetic susceptibility \mbox{$\chi_{m_Q}=(\langle m_Q^2\rangle - \langle m_Q \rangle ^2)/(TN)$},
and specific heat \mbox{$c=(\langle E^2 \rangle - \langle E \rangle^2)/(T^2N)$} (\mbox{$E=\langle \hat{H} \rangle $}).
Because antiferromagnetic interactions (\mbox{$J>0$}) are studied, staggered magnetization $m_Q$ is an order parameter in the model considered, which is calculated as a difference of magnetization of sublattices $A$ and $B$ (\mbox{$m_\alpha=(2/N)\sum_{i\in\alpha}{\langle\hat{s}_i^z\rangle}$}, \mbox{$\alpha=A,B$}).
In the antiferromagnetic (AF) phase staggered magnetization is non-zero (\mbox{$m_Q\neq0$}), whereas in the non-ordered (NO) phase \mbox{$m_Q=0$}.

\section{Results and discussion ($U/(4J)=1$)}

\begin{figure}
 \centering
 \includegraphics[width=\szerokosc\textwidth]{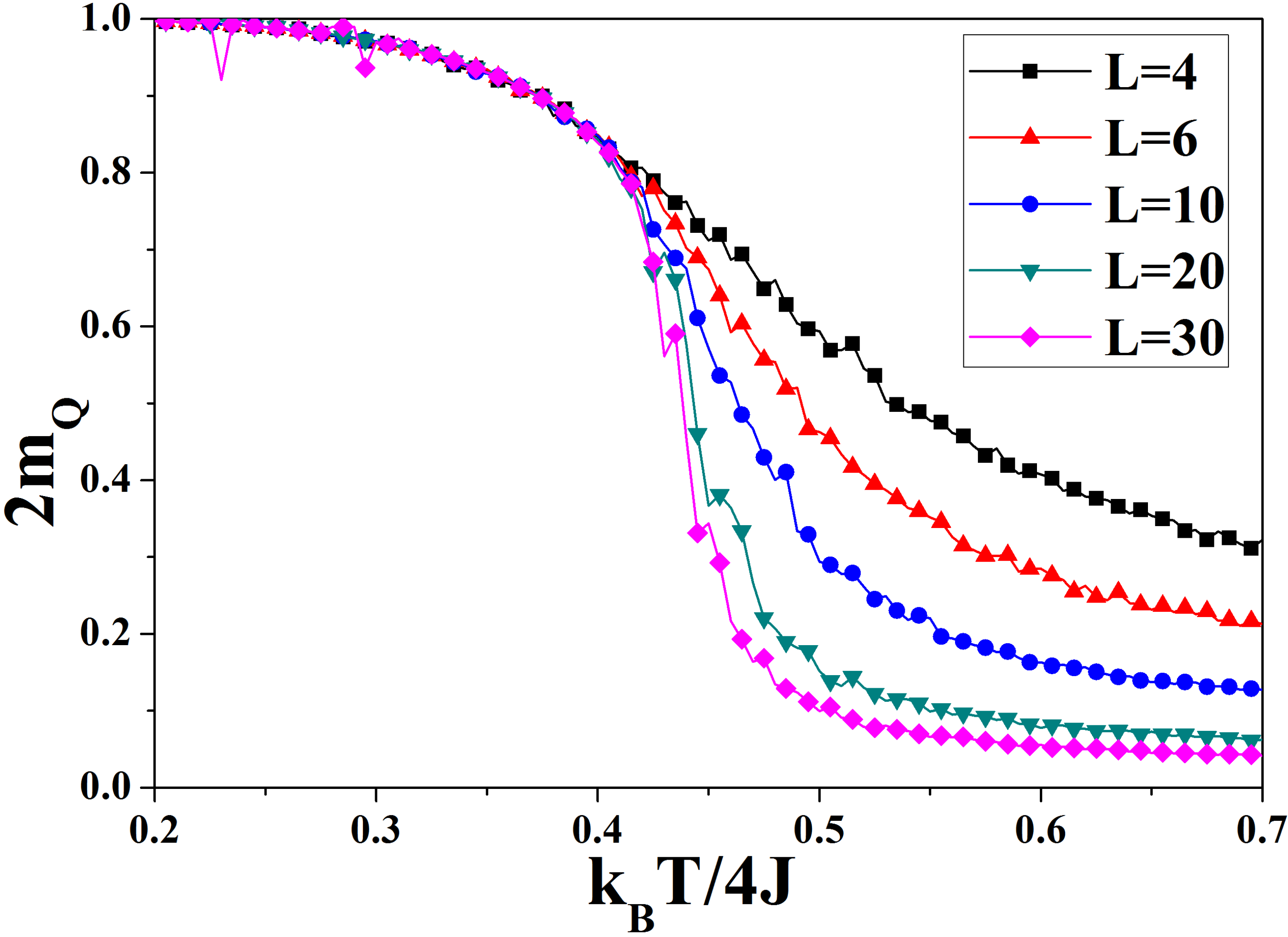}
 \caption{Magnetization $m_Q$ as a~function of temperature $k_BT/4J$  for different system sizes $L=4,\ 6,\, 10,\ 20,\ 30$ (as labelled) for \mbox{$U/4J=1$} and \mbox{$\bar{\mu}/4J=-0.22$}.}
 \label{pic:magSize}
\end{figure}

The transitions in finite systems are not sharp and the finite-size effect on the order parameter $m_Q$ is observed in the results of the MC simulations. In the NO phase $m_Q$ is larger than zero (\mbox{$m_Q\neq0$}) even above the \mbox{AF--NO} transition temperature.
The temperature dependence of magnetization $m_Q$ for different SQ lattice sizes is shown in Fig.~\ref{pic:magSize}.
While a~change from \mbox{$L=10$} to \mbox{$L=20$} yields an essential change in the results, a~further increase of $L$ does not make the transition sharper and greatly increases simulation time.
Thus system size of \mbox{$N=L\times L = 400$} has been chosen and all further results are for \mbox{$L=20$} unless said otherwise.

\begin{figure}
 \centering
 \includegraphics[width=\szerokosc\textwidth]{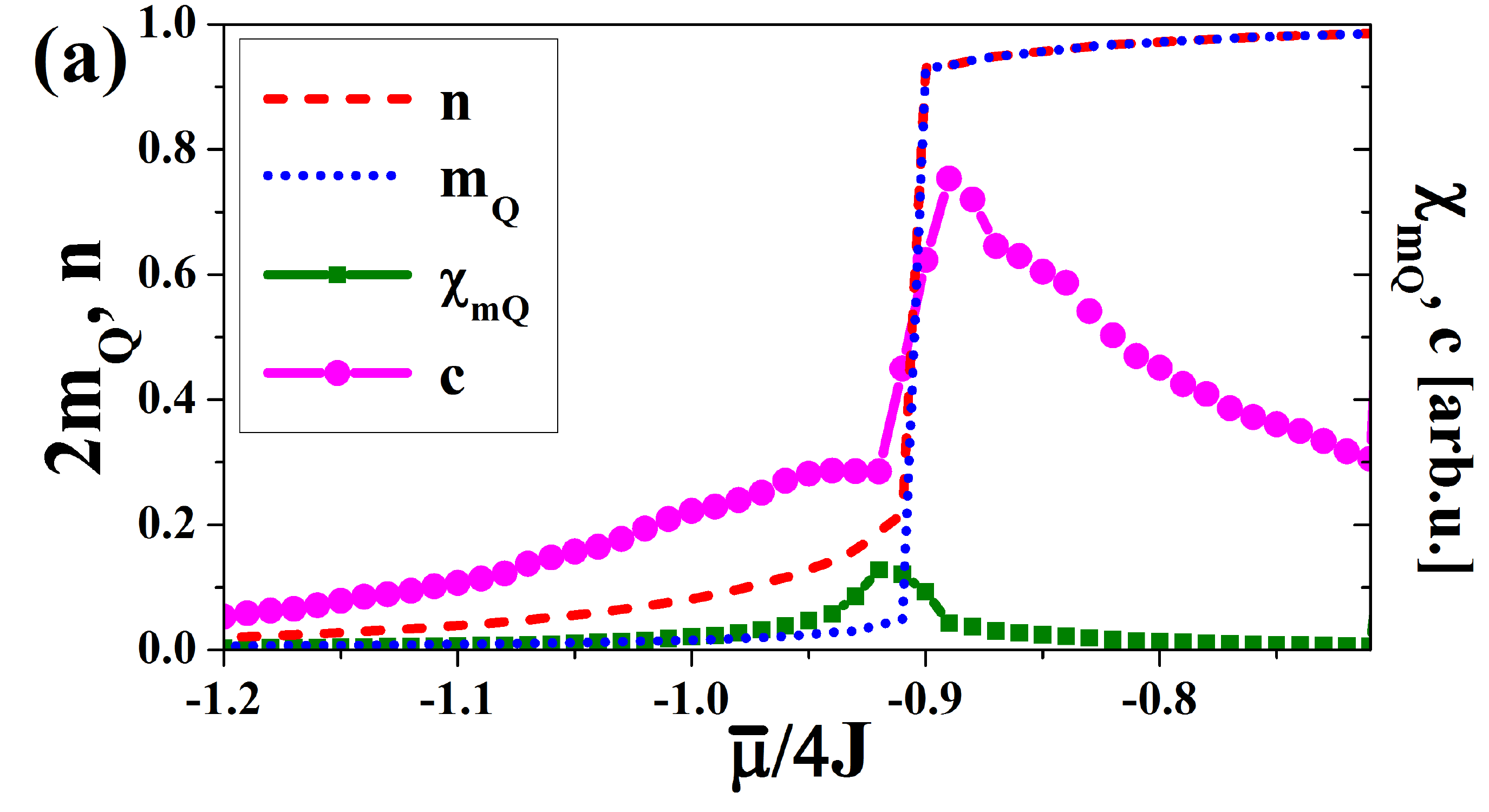}
 \includegraphics[width=\szerokosc\textwidth]{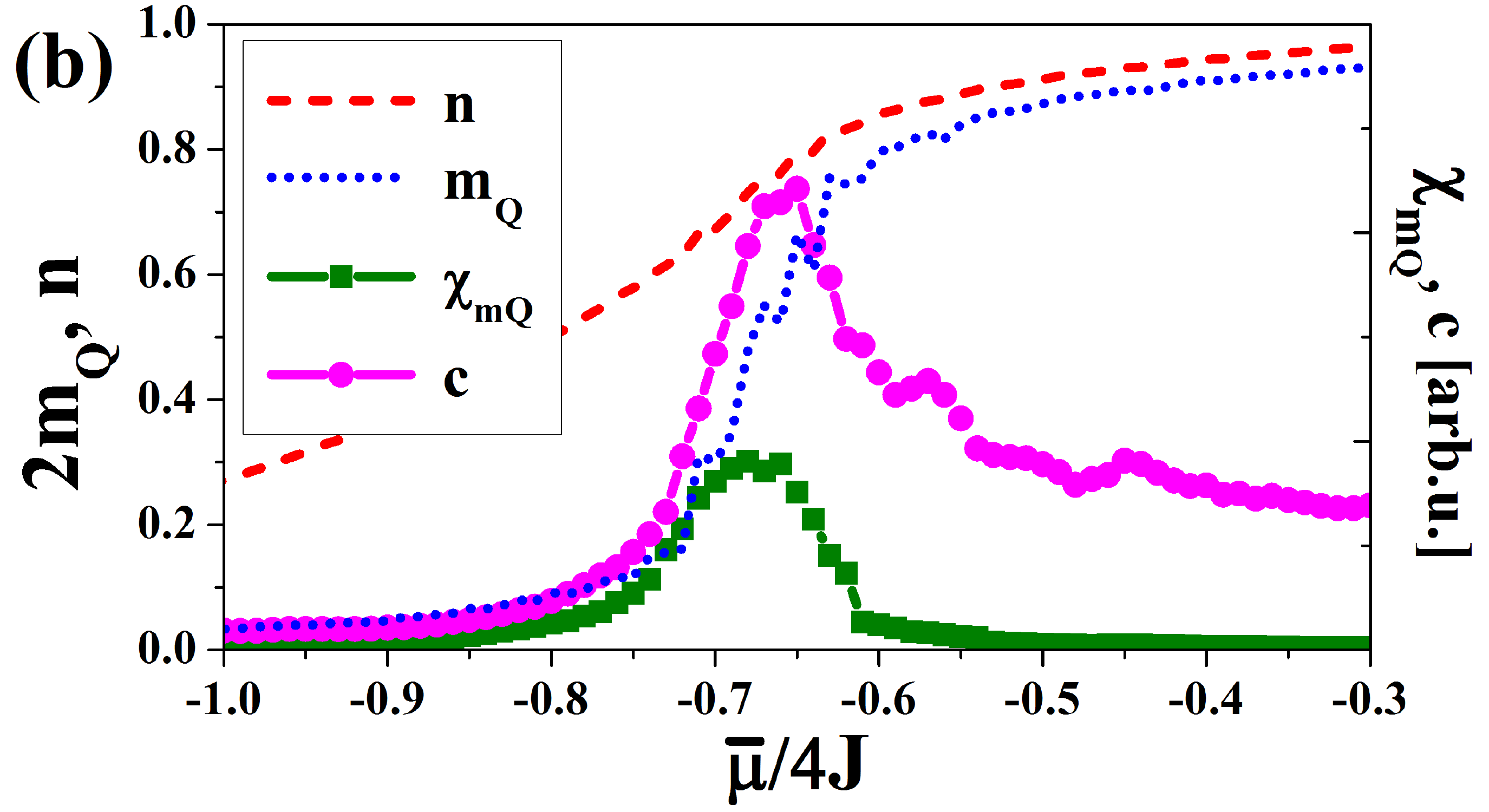}
 \includegraphics[width=\szerokosc\textwidth]{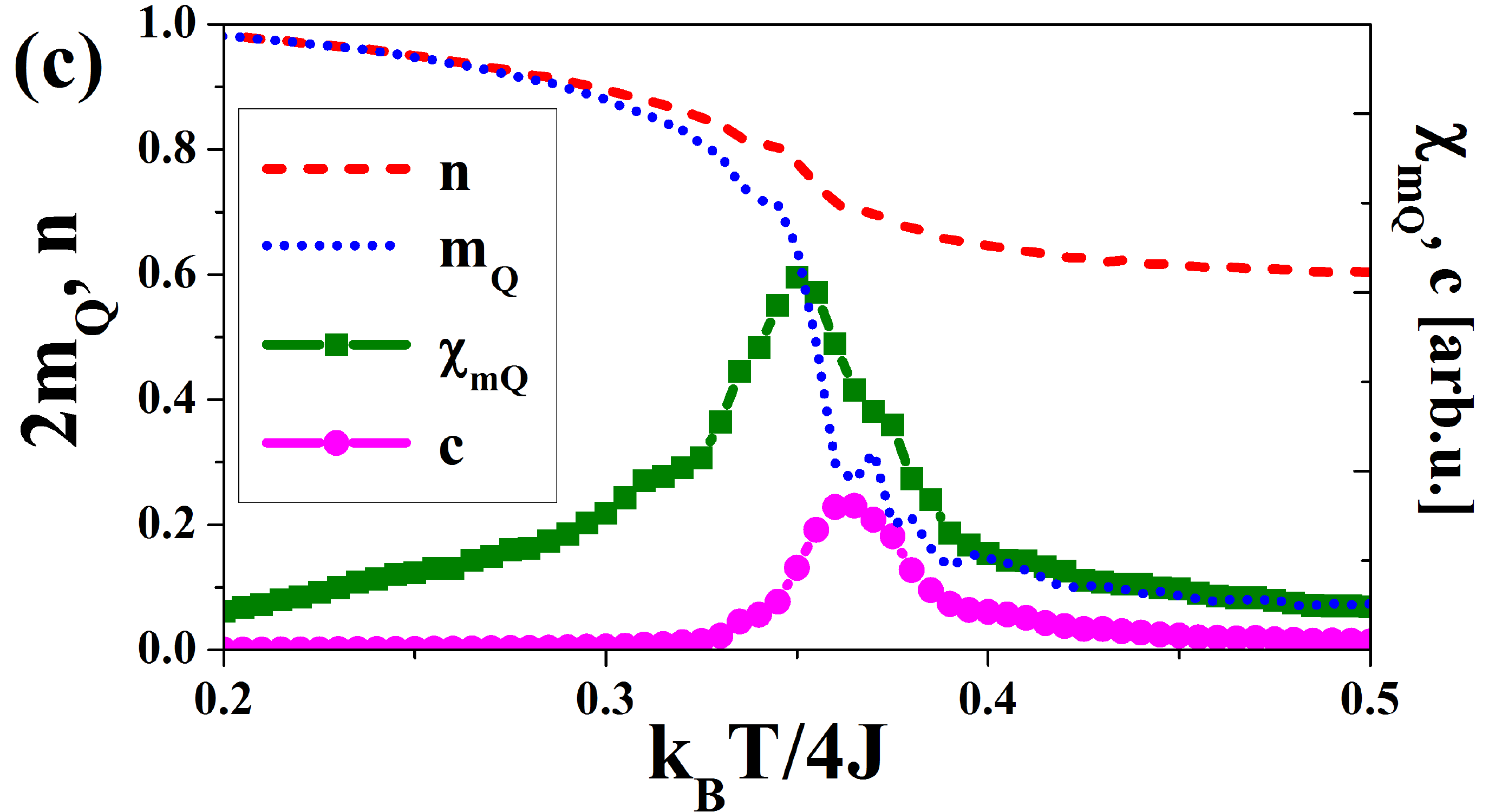}
 \includegraphics[width=\szerokosc\textwidth]{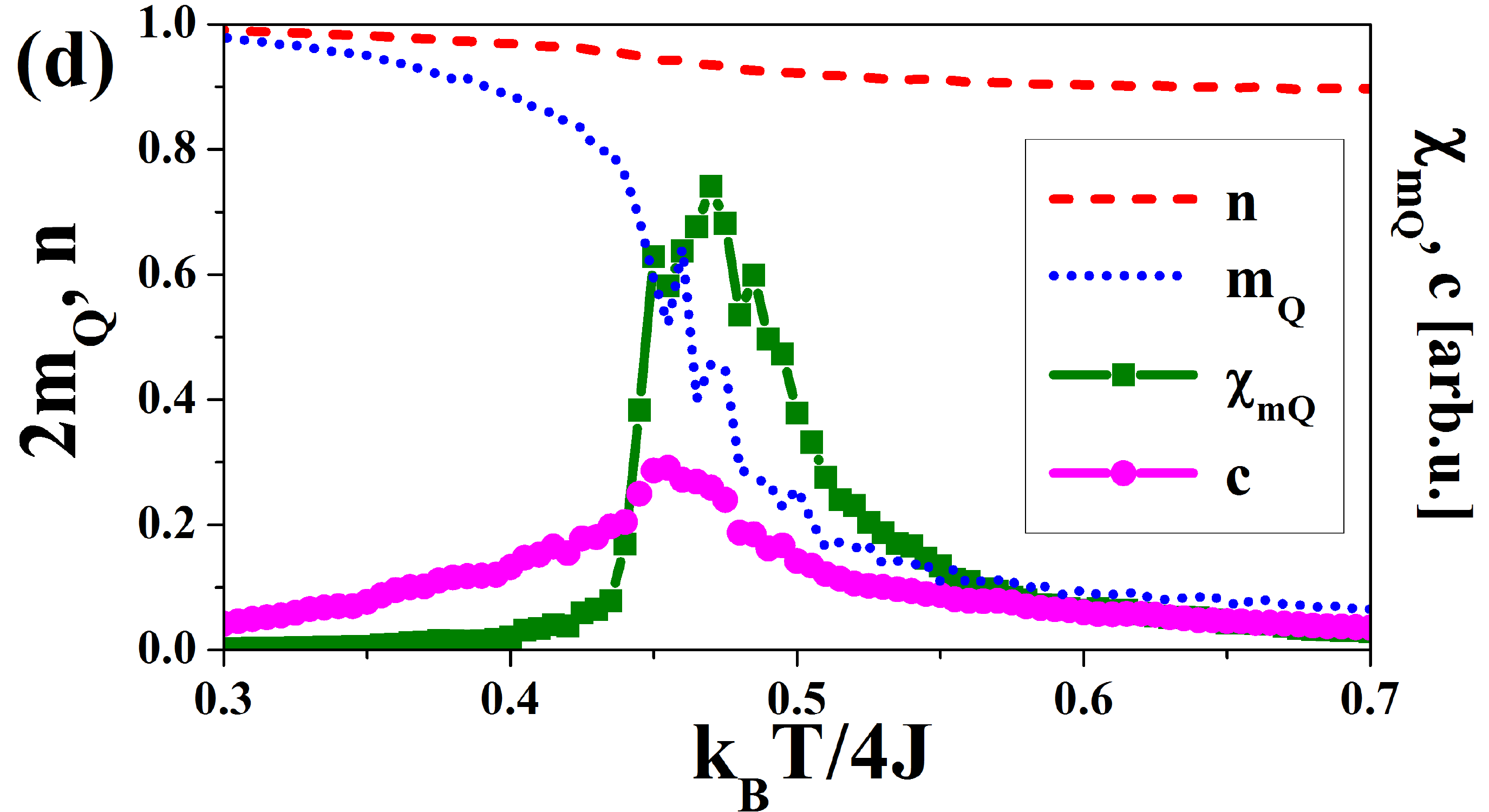}
\caption{%
Electron concentration $n$ (red), magnetization $m$ (blue), magnetic susceptibility $\chi_{m_Q}$ (green) and specific heat $c$ (violet) as a function of chemical potential $\bar\mu/4J$ for \mbox{$k_BT/4J=0.18$} (a) and \mbox{$k_BT/4J=0.36$} (b); and as a function of temperature $k_BT/4J$ for \mbox{$\bar\mu/4J=-0.67$} (c) and  \mbox{$\bar\mu/4J=-0.18$} (d) (all for \mbox{$U/4J=1$} and \mbox{$L=20$}).}
 \label{pic:properties}
\end{figure}

For a given set of model parameters temperature and chemical potential dependencies of thermodynamic properties ($n$, $m_Q$, $\chi_{m_Q}$, and $c$) have been obtained as illustrated on Fig.~\ref{pic:properties}. A~location of critical points is done by analysis of  $m_Q$,  $\chi_{m_Q}$ and $c$. The traversal of the boundary between two phases (AF and NO phases) is usually connected with a~substantial change of magnetization $m_Q$.
However, because the finite size effects is observed in the dependence of $m_Q$, a~more precise location of the critical point (i.e. AF--NO transition) is determined by the discontinuity of magnetic susceptibility $\chi_{m_Q}$ as well as a~peak in $c$.

\begin{figure}
 \centering
 \includegraphics[width=\szerokosc\textwidth]{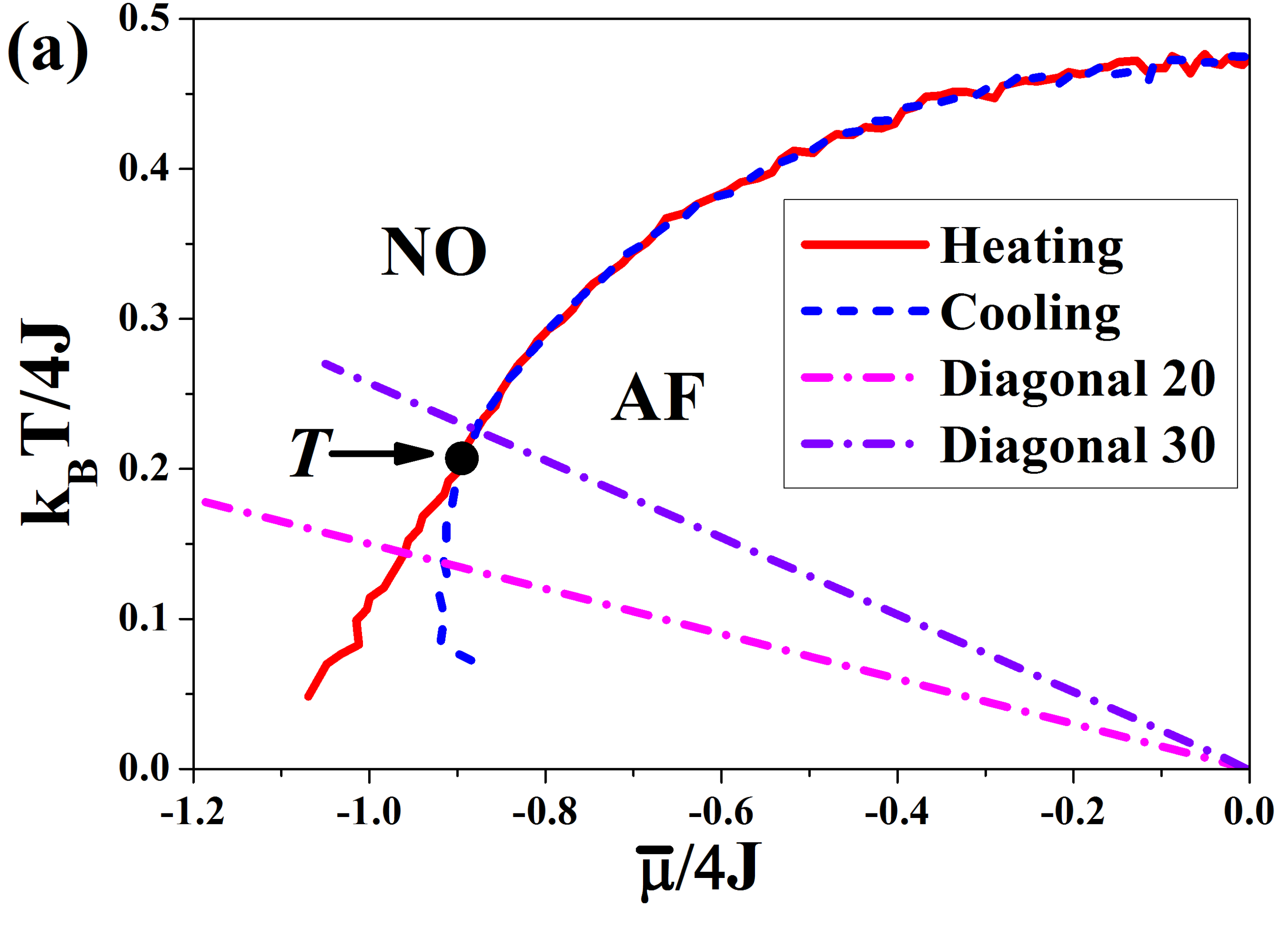}\\
 \includegraphics[width=\szerokosc\textwidth]{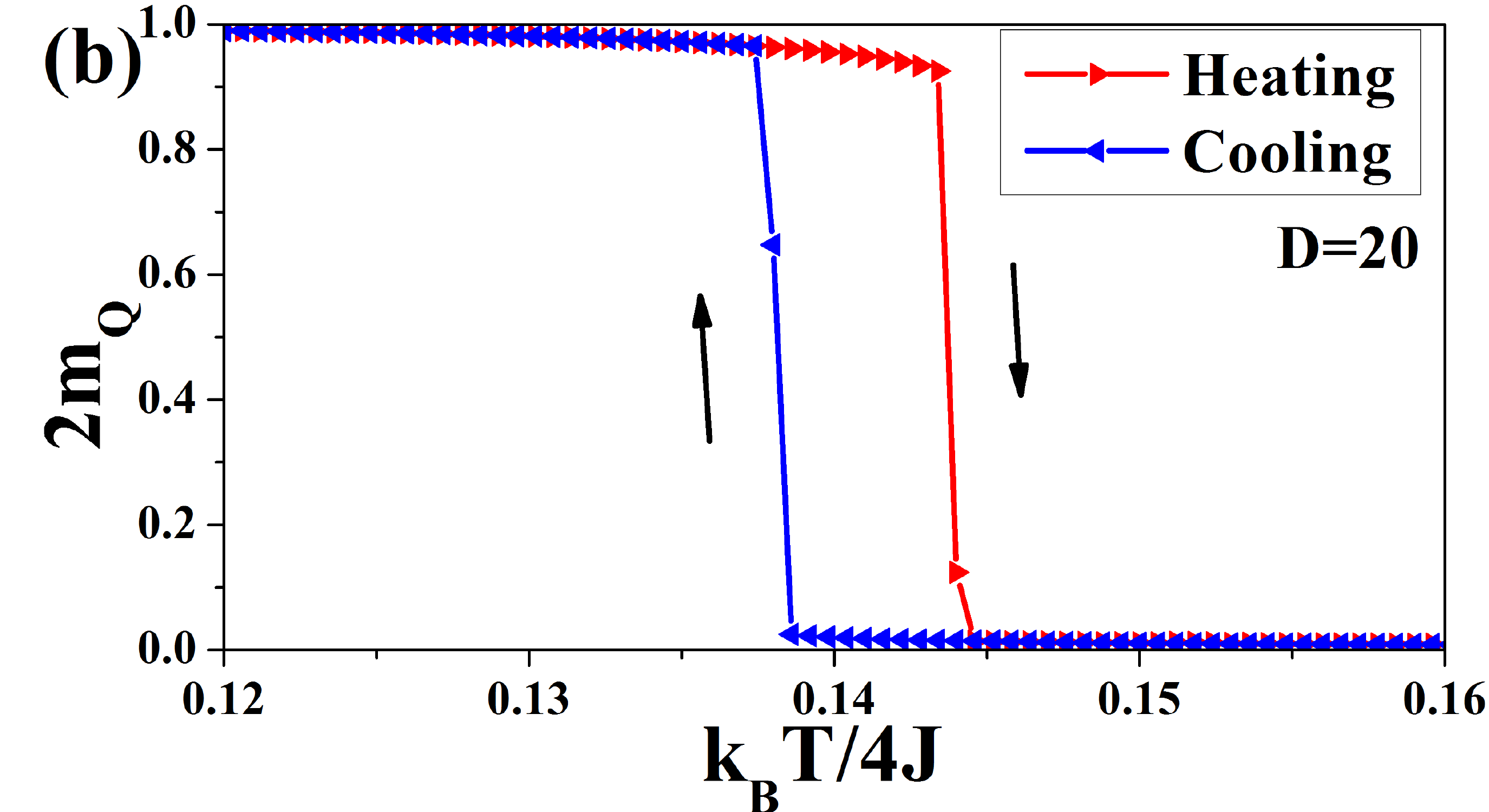}
 \includegraphics[width=\szerokosc\textwidth]{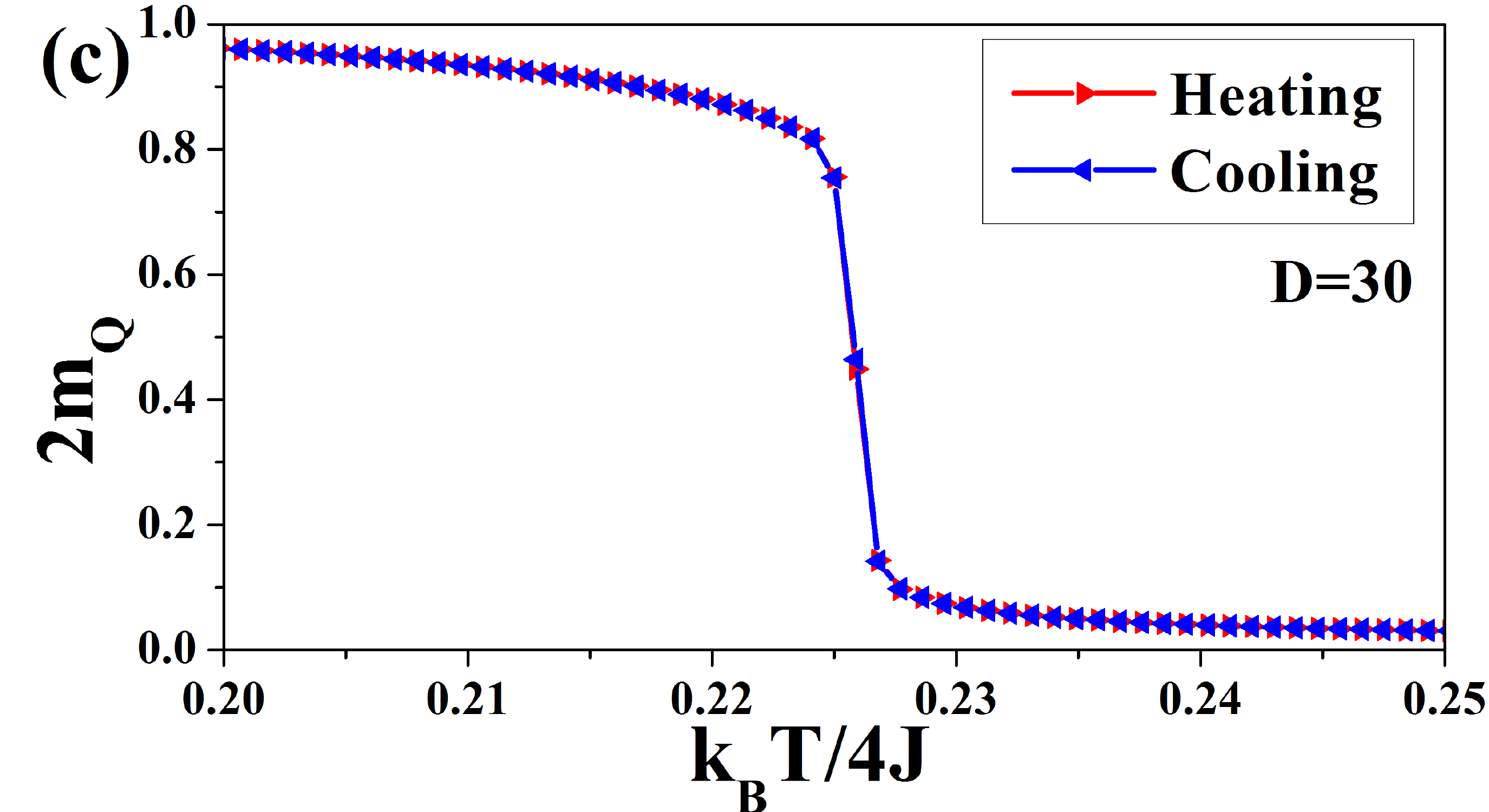}
 \caption{(a) The $k_BT/4J$ vs. $\bar{\mu}/4J$ phase diagram for \mbox{$U/4J=1$}. $\mathbf{T}$~denotes a~tricritical point. Dashed-dotted lines indicate diagonals labeled as $20$ and $30$. (b), (c) Staggered magnetization $m_Q$ for heating (red) and cooling (blue) runs  for diagonal index \mbox{$D=20$} (first-order transition) (b) and \mbox{$D=30$} (second-order transition) (c) as a function of temperature $k_BT/4J$ (all for \mbox{$U/4J=1$} and \mbox{$L=20$}).}
  \label{pic:propertiesDiagonal}
\end{figure}

Instead of running simulations with fixed $k_BT/4J$ or $\bar{\mu}/4J$, the simulations with both these thermodynamic parameters changing have been also performed. In such an approach the system moves diagonally in \mbox{$(\bar\mu,T)$} plane, approaching the phase transition area almost perpendicular, which allows for more precise determination of the \mbox{AF--NO} transition temperature.
We introduce definition of a diagonal ($D$), where a~diagonal labelled as $0$ is  perpendicular to \mbox{$k_BT/4J$}--axis and a~diagonal labeled as $100$ is parallel to it. For each step on the diagonal $D$ the parameters change as \mbox{$\Delta T_{D}=D\Delta T /100$}, \mbox{$\Delta \bar{\mu}_{D}=(100-D)\Delta \bar{\mu}/100$}, where $D$ is diagonal number, and $\Delta T$, $\Delta \bar{\mu}$ correspond to temperature ($k_BT/4J$) and chemical potential ($\bar{\mu}/4J$) fixed steps on the respective axes.
The final $k_BT/4J$ vs. $\bar{\mu}/4J$ phase diagram for \mbox{$U/4J=1$} is shown in Fig.~\ref{pic:propertiesDiagonal}(a). In Fig.~\ref{pic:propertiesDiagonal}(a) there are also indicated the diagonals with indexes \mbox{$D=20$} and \mbox{$D=30$}.
Staggered magnetization $m_Q$ for heating with removal of particles and cooling with addition of particles for diagonal index $20$ and $30$  as a function of temperature for \mbox{$U/4J=1$} are shown in Figs.~\ref{pic:propertiesDiagonal}(b) and (c), respectively.

The \mbox{AF--NO} transition  temperatures  increase monotonically with  decreasing $|\bar{\mu}|$.  The maximum of the transition temperature is located at \mbox{$\bar{\mu}=0$} (\mbox{$n=1$}). For  \mbox{$U/4J=1$} and \mbox{$n=1$} the critical temperature is equal to \mbox{$k_BT/4J\simeq0.47$} (\mbox{$L=20$}).

To determine the location of the tricritical point $\mathbf{T}$ two simulation runs have been done for each diagonal, one starting at \mbox{$(0,0)$} in the \mbox{$\bar\mu/4J$--$k_BT/4J$} plane and the other starting at the maximum point for the given diagonal and descending to \mbox{$(0,0)$}. This corresponds to heating and cooling processes, respectively. A~position of the $\mathbf{T}$ point can be estimated by comparing magnetization data at the point  of the phase transition for those two simulations. For second-order phase transition both magnetization curves should be almost identical (Fig.~\ref{pic:propertiesDiagonal}(c)), while first-order phase transition is characterized by the hysteresis (Fig.~\ref{pic:propertiesDiagonal}(b)). Thus, a~point, where the hysteresis is collapsing into single curve, is a~tricritical point. For \mbox{$U/4J=1$} (\mbox{$L=20$}) the $\mathbf{T}$ point is located at \mbox{$k_BT/4J=0.205\pm0.005$} and \mbox{$\bar\mu/4J=-0.895\pm0.007$} (\mbox{$n=0.52\pm0.2$}).

\begin{figure}
 \includegraphics[width=\szerokosc\textwidth]{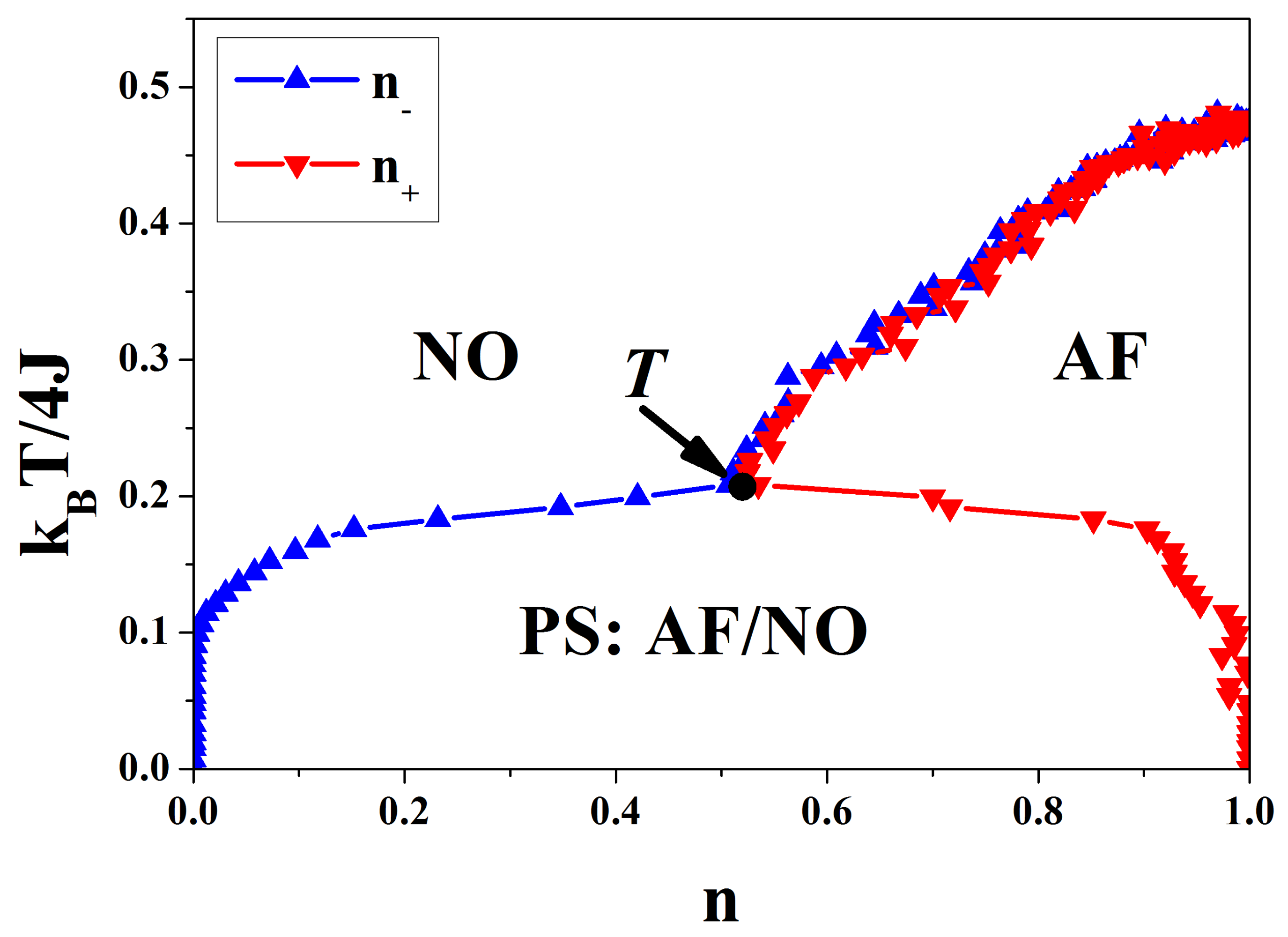}
 \caption{The $k_BT/4J$ vs. $n$ phase diagram for \mbox{$U/4J=1$} (and \mbox{$L=20$}). $\mathbf{T}$~indicates a~tricritical point.}
 \label{pic:phase shift}
\end{figure}

With simulations done for fixed $\bar{\mu}$, it is  possible to obtain phase diagrams as a function of $n$ by determining electron density above ($n_{-}$) and below ($n_{+}$) the \mbox{AF--NO} phase transition (for fixed $\bar{\mu}$).
The first-order \mbox{AF--NO} boundary for fixed $\bar{\mu}$ splits into two boundaries (i.e. \mbox{PS--AF} and \mbox{PS--NO}) for fixed $n$.
For \mbox{$U/4J=1$} the $k_BT/4J$~vs.~$n$ phase diagram is presented in Fig.~\ref{pic:phase shift}. For temperatures above the tricritical point $\mathbf{T}$, the AF and NO phases are  separated by second-order line. At lower temperatures, below this point, there is a~phase separated (PS: AF/NO) state. The PS state is a~coexistence of two (AF and NO) homogeneous phases.

Comparing the results presented in this paper with preliminary MC simulations from previous work \cite{MKPR2012} of our group, the general improvement of their  quality is clearly seen. Increasing the SQ lattice size as well as simulating along diagonals gives more accurate values of critical temperatures. Moreover, performing heating and cooling simulation runs yields much more precise location of the tricritical point.

\section{Final comments}

The results presented in this paper are in a~good qualitative agreement with those obtained within the variational  approach  (VA) involving the mean-field  approximation for intersite interactions \cite{KKR2010,MKPR2012}, which is exact in \mbox{$d\rightarrow+\infty$} (\mbox{$L\rightarrow\infty$}, \mbox{$z\rightarrow\infty$}).
Our MC simulations have been performed for \mbox{$d=2$} SQ lattice. In this case the VA is much less reliable. It largely overestimates critical temperatures, e.g. \mbox{$k_BT/4J\simeq0.68$} at \mbox{$n=1$}, and yields a~quite different location of the $\mathbf{T}$ point: \mbox{$k_BT/4J=1/3$}, \mbox{$\bar{\mu}/4J\simeq-0.96$}, \mbox{$n\simeq0.34$} (for \mbox{$U/4J=1$}) \cite{KKR2010,MKPR2012}.
While finite-size effects do not pose a big problem, long thermalization time at low temperatures prevents from obtaining results near the ground state. At such low temperatures the probability of an electron to change its state is minimal, so the system has little chance of escaping false energy minima. A~solution of this problem is running simulations for a very long time, or collecting results from different ``runs'' for various starting states.

It is important to mention that in the absence of an external magnetic field the antiferromagnetic (\mbox{$J > 0$}) interactions are simply mapped onto the ferromagnetic ones (\mbox{$J < 0$}) by redefining the spin direction in one sublattice in alternate lattices decomposed into two interpenetrating sublattices. Thus, our results obtained in this paper are still valid for \mbox{$J<0$} if \mbox{$m_Q\rightarrow m= (m_A+m_B)/2$} and \mbox{$J\rightarrow|J|$}.

The analysis of effects of finite band-width (\mbox{$t\neq 0 $}) is a~very important problem. However, because of the complexity of such model only few results are known beyond weak coupling regime or away half-filling \cite{MRR1990,JM2000,DJSZ2004,DJSTZ2006,CzR2006,CzR2006a,CzR2007}. For instance, the presence of the hopping term \mbox{$\sum_{i,j,\sigma}{t_{ij}\hat{c}^{+}_{i\sigma}\hat{c}_{j\sigma}}$}   breaks a~symmetry between \mbox{$J<0$} and \mbox{$J>0$} cases \cite{JM2000,DJSZ2004,CzR2006}. The detailed analysis and discussion on this topic is left for future study.

The competition between magnetism and  superconductivity \cite{RP1993,K2012,KRM2012,K2014,KR2013,K2013,K2014b} in atomic limit of the extended Hubbard models is a~very interesting topic. Some results concerning the interplay of magnetic interactions with the pair hopping term  have been presented in \cite{K2012}. Moreover the interplay between various magnetic and charge orderings \cite{MRC1984,MM2008,KR2010,KR2011a,KR2011b,KR2012,MPS2013B} has been also analysed~\cite{KKR2012,MPS2011}.

\begin{acknowledgments}
S.M. and K.K. thank the European Commission and the Ministry of Science and Higher Education (Poland) for the partial financial support from the European Social Fund---Operational Programme ``Human Capital''---POKL.04.01.01-00-133/09-00---``\textit{Proinnowacyjne kszta\l{}cenie, kompetentna kadra, absolwenci przysz\l{}o\'sci}''.
K.K. and S.R. thank  National Science Centre (NCN, Poland) for the financial support as a~research project under grant No. DEC-2011/01/N/ST3/00413 and as a~doctoral scholarship No. DEC-2013/08/T/ST3/00012.
K.K. thanks also the Foundation of Adam Mickiewicz University in Pozna\'n for the support from its scholarship programme.
\end{acknowledgments}


\end{document}